\documentclass{ptapap}
\usepackage[]{natbib}

\author{Stefan G\"ossl}[Stefan.Goessl@student.uibk.ac.at,UIBK]
\author{Konstanze Zwintz}[konstanze.zwintz@uibk.ac.at,UIBK]
\author{Rainer Kuschnig}[rainer.kuschnig@univie.ac.at,TUG]
\affil[UIBK]{Universit\"at Innsbruck, Institute for Astro- and Particle Physics, \\
  Technikerstrasse 25/8, A-6020 Innsbruck, Austria}
\affil[TUG]{Graz University of Technology, Institute for Communication Networks and Satellite Communication, 
Inffeldgasse 12, A - 8010 Graz}

\title{43 Cygni observed with BRITE-Constellation}

\begin{document}

\maketitle

\begin{abstract}

The $\gamma$ Doradus star 43 Cygni was observed by BRITE-Toronto for 156 days in 2015. From this data set we identified 37 pulsation frequencies which show a regular period spacing pattern that will be subject of a more detailed asteroseismic analysis.

\end{abstract}

\section{Introduction}

Only $\sim$12\% of the targets observed by BRITE-Constellation have spectral types F. Among those are a few $\gamma$ Doradus type pulsators which show gravity ($g$) mode pulsations excited by the convective flux block mechanism \citep{guzik00}. The observed pulsation periods range from 0.3 to 3 days and are predicted to be equidistantly spaced with periodic dips in the spacing pattern if a chemical gradient is present at the edge of the convective core \citep{miglio08}. Only recently period spacing patterns were discovered in photometric time series of $\gamma$ Doradus stars \citep[e.g.,][]{vanreeth15aa} observed with the Kepler space telescope \citep{koch10}.

43 Cygni (HD 195068/9, V 2121 Cygni, HR 7828)  was originally identified as possible RR Lyrae star from HIPPARCOS Photometry \citep{perryman97}. \citet{fekel03} found asymmetries in the line profiles and suggested that they could be either from binarity or from pulsations. The identification as a (single) $\gamma$ Doradus pulsator was finally conducted by \citet{henry05} who detected three main frequencies at f1 = 1.25054\,d$^{-1}$, f2 = 1.29843\,d$^{-1}$ and f3 = 0.96553\,d$^{-1}$, using photometric time series in the Johnson $B$ and $V$ filters.

Using 230 spectra obtained over two years, \citet{jankov06} studied the line profile variations, detected an additional frequency and attempted a first spectroscopic mode identification. The authors reported a frequency at 1.61\,d$^{-1}$ to be an $\ell$ = 5 ($\pm$1) and m = 4 ($\pm$ 1) and the frequency at 1.25\,d$^{-1}$ \citep[which is f1 in ][]{henry05} to be an $\ell$ = 4 ($\pm$1) and m = 3 ($\pm$ 1) mode.

\citet{gerbaldi07} identified 43 Cyg as single star from TD1 UV observations, and \citet{cuypers09} confirmed the previously identified pulsation frequencies from photometric time series in the Geneva system.

43 Cygni has an effective Temperature, $T_{\rm eff}$, of 7300 $\pm$ 250 K and a log\,$g$ of 4.35 $\pm$ 0.14 \citep{david15}. Its projected rotational velocity, $v$\,sin\,$i$, is 44 kms$^{-1}$ \citep{fekel03}.
As 43 Cygni has a $V$ magnitude of 5.74, it belongs to the faintest stars observable with BRITE-Constellation.

\section{BRITE-Constellation Observations}
\begin{figure}[htb]
\centering
\includegraphics[width=0.8\textwidth]{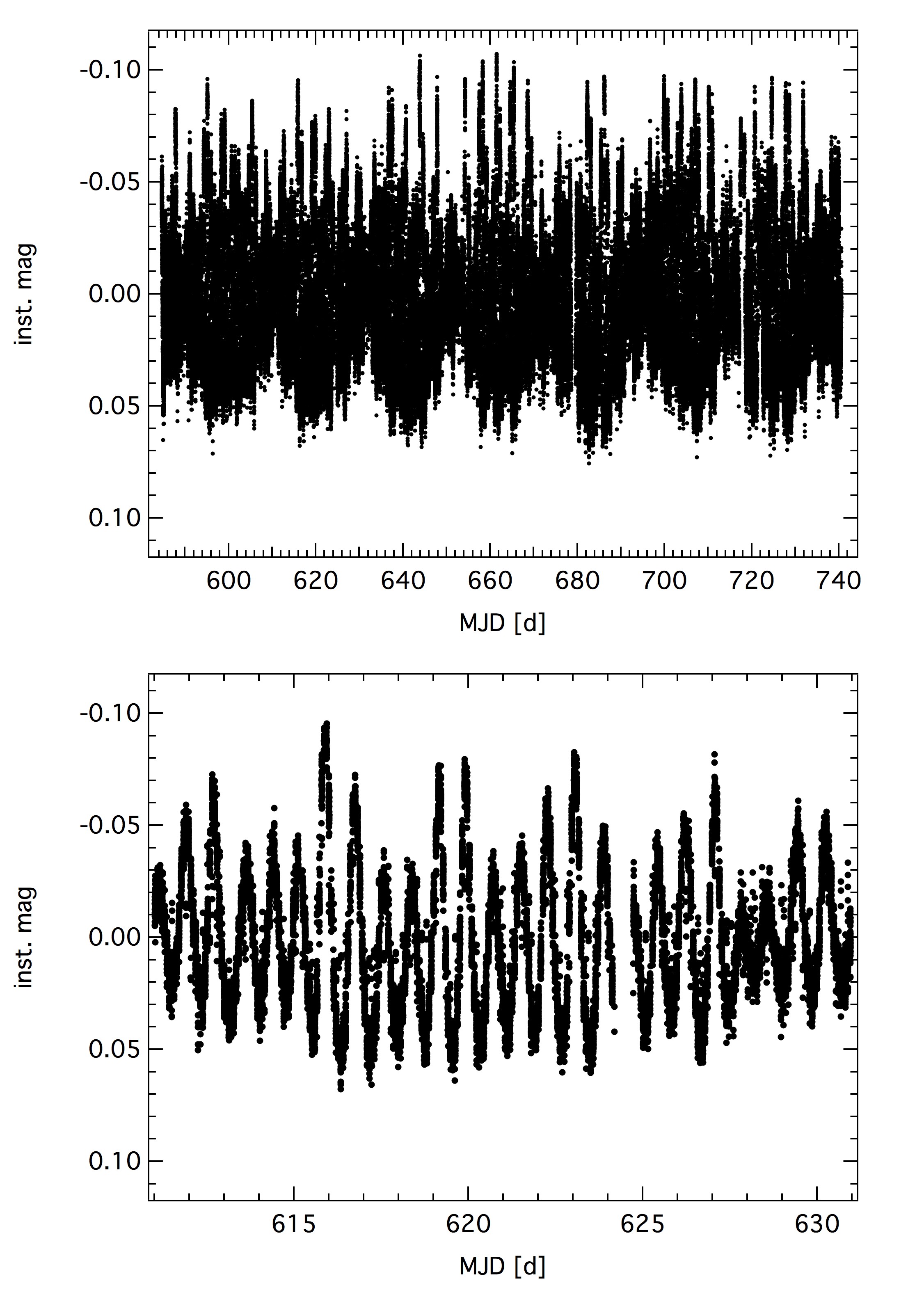}
\caption{BRITE-Toronto light curve of 43 Cygni (top panel) and zoom into a 20-day subset (bottom panel).}
\label{43cyg_lc}
\end{figure}

BRITE-Constellation obtained observations of the Cygnus-II field from June 1 to November 25, 2015. 43 Cygni was observed for only 13 days with Lem in the blue filter and for $\sim$156 days with BRITE-Toronto in the red filter, both in chopping mode \citep{pablo16}. As the blue data set from Lem did not have sufficient quality for an asteroseismic analysis, we omitted it from our further investigation. The complete light curve of 43 Cygni obtained with BRITE-Toronto is shown in the top panel of Figure \ref{43cyg_lc}; the bottom panel shows a zoom into a 20-day subset.

\section{Data Reduction}

Data reduction was performed using the software developed by M. Kondrak and described in detail in these proceedings.
Frequency analysis was conducted using the program {\sc Period04} \citep{period04} which combines Fourier and least-squares algorithms. Frequencies were then prewhitened and considered to be significant if their amplitudes exceeded four times the local noise level in the amplitude spectrum \citep{breger93,kuschnig97}.

\begin{figure}[htb]
\centering
\includegraphics[width=0.8\textwidth]{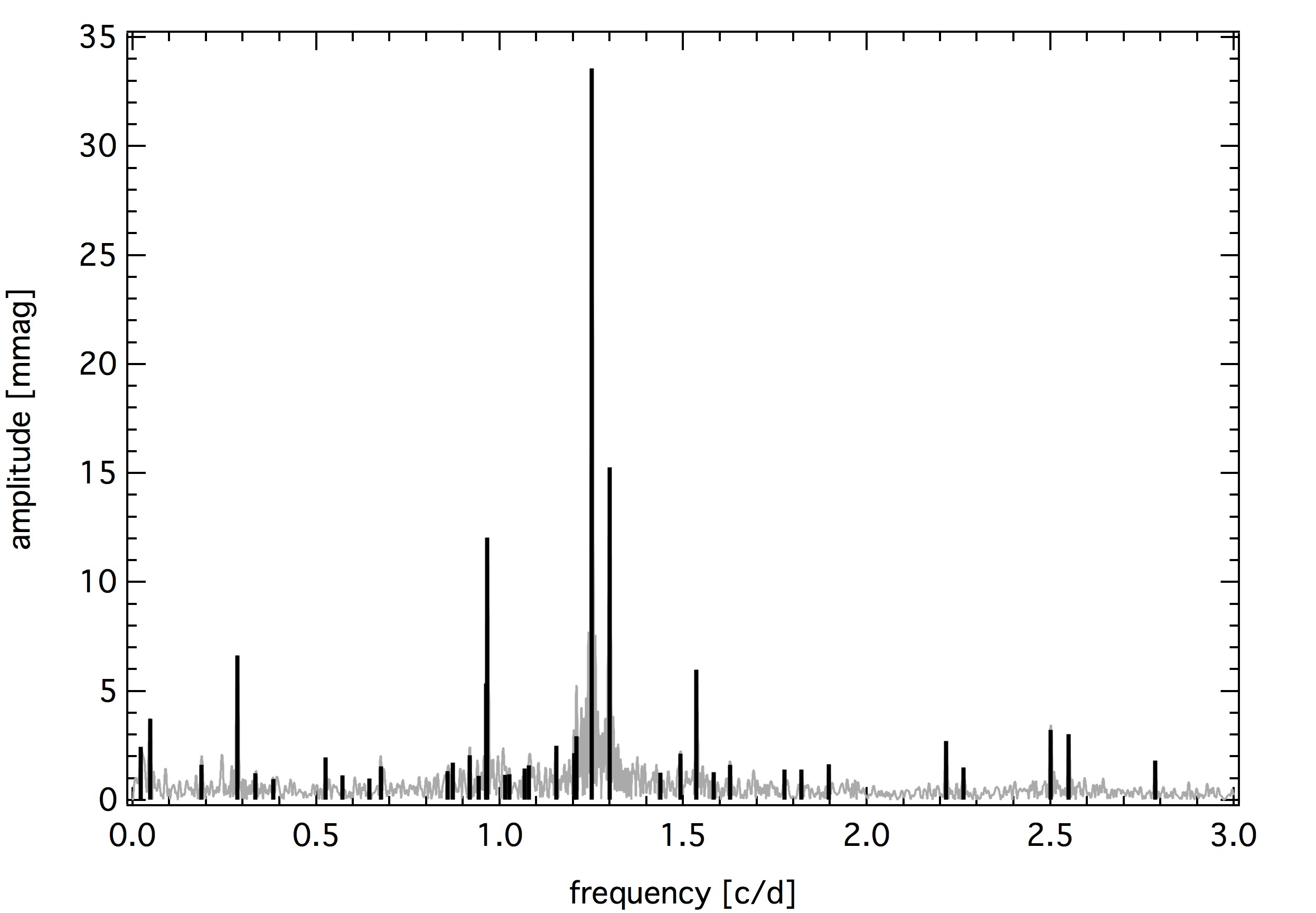}
\caption{BRITE-Toronto light curve of 43 Cygni (top panel) and zoom into a 20-day subset (bottom panel).}
\label{ampspec}
\end{figure}

In total we detected 37 intrinsic frequencies in a range from 0 to 2.5\,d$^{-1}$ that are caused by $\gamma$ Doradus type pulsations (see Figure \ref{ampspec}). We can also confirm the previously identified frequencies at 1.25054\,d$^{-1}$, 1.29843\,d$^{-1}$, 0.96553\,d$^{-1}$ and 1.61\,d$^{-1}$ reported by \citet{henry05} and \citet{jankov06}. The 37 corresponding periods show a clear regular spacing pattern that is currently being investigated asteroseismically and will lead to an asteroseismic mode identification. 

\section{Conclusions}

The 156 days long photometric time series obtained with BRITE-Toronto revealed 37 pulsation frequencies for 43 Cygni which show an equidistant period spacing pattern that will lead to an identification of the pulsation modes and the determination of the star's rotation rates. 
 
These results illustrate the potential of BRITE-Constellation observations for $\gamma$ Doradus stars: The up to 180 days long data sets observable with the BRITE nano-satellites allow to detect a significant number of pulsation frequencies and to conduct thorough asteroseismic analyses through investigation of period spacing patterns similar as it is currently only possible with the 4-years light curves from the main Kepler mission. 

\acknowledgements{
Based on data collected by the BRITE-Constellation satellite mission, built, launched and operated thanks to support from the Austrian Aeronautics and Space Agency (FFG) and the University of Vienna, the Canadian Space Agency (CSA) and the Foundation for Polish Science \& Technology (FNiTP MNiSW) and National Centre for Science (NCN).
SG acknowledges support from the project ``Asteroseismology with BRITE-Constellation'' (Nachwuchsf\"orderung der Universit\"at Innsbruck 2015, PI: K. Zwintz). 
K.Z. acknowledges support by the Austrian Fonds zur F\"orderung der wissenschaftlichen Forschung (FWF, project V431-NBL).

}

\bibliographystyle{ptapap}
\bibliography{SGoessl}

\end{document}